\documentclass{article}
\usepackage[utf8]{inputenc}
\usepackage{dsfont}
\usepackage{amssymb}
\usepackage{amsmath}
\usepackage{amsthm}
\usepackage{amsfonts}
\usepackage{mathabx}
\usepackage{enumitem}
\usepackage{hyperref}
\usepackage{graphicx}
\usepackage{tabularx}
\usepackage{caption}
\usepackage{subcaption}
\usepackage[all]{xy}
\usepackage{etoolbox}
\usepackage{multirow}
\usepackage{makecell} 

\usepackage{qcircuit}

\usepackage{hyperref}

\hypersetup{
    colorlinks=true,
    linkcolor=blue
    }

\title{Layered Uploading for Quantum Convolutional Neural Networks}
\author{Grégoire Barrué, Tony Quertier, Orlane Zang}
\date{}

\begin{document}
\maketitle

\begin{abstract}
Continuing our analysis of quantum machine learning applied to our use-case of malware detection, we investigate the potential of quantum convolutional neural networks. More precisely, we propose a new architecture where data is uploaded all along the quantum circuit. This allows us to use more features from the data, hence giving to the algorithm more information, without having to increase the number of qubits that we use for the quantum circuit. This approach is motivated by the fact that we do not always have great amounts of data, and that quantum computers are currently restricted in their number of logical qubits. 
\end{abstract}

\section*{Introduction}
A lot of work has been done on parameterized quantum circuit in order to understand their capacities especially for quantum machine learning applications \cite{Cerezo_2022}. Expressiveness and generalization capabilities have been theoretically explored \cite{Abbas_2021,Larocca_2021,Sim_2019}, enlightening some specific issues to the quantum setting. One of the most important issue seems to be the Barren Plateau phenomenon \cite{Holmes_2022, McClean_2018}, where gradients vanish exponentially fast with the increase of the number of qubits, causing the failure of the algorithm's learning. In turns out that the Lie algebra is an important tool to evaluate whether a quantum neural network is subject to Barren plateaus or not \cite{ragone2023unified,Wiersema_2020}, and some initialization strategies have been tested \cite{Grant_2019}, as approaches to keep track of the data all along the circuits \cite{P_rez_Salinas_2020}. A particular alternative is also quantum convolutional neural networks (QCNNs) because it has been proved that under some hypothesis they do not exhibit Barren Plateaus \cite{Pesah_2021}. They were first introduced in \cite{cong2019quantum}, and used for instance in  \cite{Hur_2022} for classical data classification, or in \cite{Bokhan_2022} for multi-class classification. Their full understanding is still missing, but some works, as \cite{umeano2023learn}, start to deeply analyze these models, and we think that it could be a suitable solution for image classification, as is its classical equivalent.

This classification task is a big interest for us, as we want to use it for malware detection, which is a current area of research due to the ever-increasing number of attacks. In this context, researchers have turned to artificial intelligence to improve the detection of new malware  \cite{Gibert2020, Raff2020, Ucci2019}, and some techniques have been developed to learn about features extracted from binary semantic and statistical data \cite{Anderson2018}, to use language processing elements \cite{raff2018malware} or even convolutional neural networks (CNNs) \cite{marais2023ameliorations,marais2022ai}. 

In this article, we propose a new architecture for quantum convolutional networks, allowing to inject more features in the QCNN, and thus more information about the data, without increasing its size. We compare this architecture with a standard QCNN on two different datasets, first the well-known MNIST dataset and second a dataset composed of images corresponding to malware and benign PE files. Our motivation here is to prove that there are solutions to use more information in cases where great amounts of data are not available, or computations resources are limited. Thus we use relatively few data and models with little numbers of qubits.

This article is organized as follows. In Section \ref{sec:Preprocessing} we detail the datasets that we use, especially the dataset containing malware files. We explain the preprocessing that we used, named Grayscale method, which allows to turn a PE file into an image, in order to perform image classification. Then Section \ref{sec:descriptionQCNN} details the implementation of the standard quantum convolutional neural network that we use for our tests, and states the absence of Barren plateaus for this model. Finally Section \ref{sec:Reuploading_circuit} proposes our new architecture, layered uploading, and gathers the results of our experiments to highlight the efficiency of this method.

% Malicious software detection has become an important topic in business, as well as an important area of research due to the ever-increasing number of successful attacks using malware. With the recent advances in Artificial Intelligence (AI), cybersecurity researchers are shifting their attention to Machine Learning (ML) and Deep Learning (DL) methods to improve malicious files detection \cite{Raff2020, Ucci2019}, and they have been incredibly creative in data preprocessing. In fact, this part is essential now that the learning algorithms are already extremely powerful.

% In \cite{Anderson2018}, Anderson et al. trained a feature-based malware detection model using a non-optimized LightGBM algorithm, in \cite{Nataraj2011} Nataraj et al. use k-nearest neighbors algorithm on image-based malware whereas Raff et al. \cite{Raff2017} introduced MalConv, a featureless deep learning classifier using a dense neural network processing raw bytes of entire executable files.
% Image-based malware detection is a challenge in both classical and quantum computing, but for different reasons. In classical machine learning, one limitation to obtaining results as good as with standard static features \cite{https://doi.org/10.48550/arxiv.2107.11100} is the number of training data. Because malware images are much more complex and less representative than standard images, convolutional networks need a lot of images to extract information. However, computing resources are not a problem for two-channel images of size $64 \times 64$.

\section{Preprocessings and datasets}
\label{sec:Preprocessing}

Our principal interest is to study malware detection. In this aim we rely on two different datasets,  Bodmas \cite{Yang} and PEMachineLearning \cite{web1}. They all contain malicious and benign Portable Executable (PE) files, distribution and format are summarized in Table \ref{tab:dataset}.

\begin{table}[!ht]
    \centering
     \caption{Distribution and format of each dataset}
    \label{tab:dataset}
    \begin{tabular}{|c|c|c|}
        \hline
         Dataset & Malicious files & Benign files  \\

        \hline
        Bodmas &57,293 & 77,142\\

        \hline
        PEMachine Learning & 114,737 & 86,812 \\
        \hline
    \end{tabular}
\end{table}

Bodmas \cite{Yang} provided us  a dataset composed of $134,435$ binary files with pre-extracted features together with the $57,293$ malicious files in raw PE format. These files have been collected during one year between August 2019 and September 2020 and labeled: authors indicate the category each file belongs to. The other dataset, PEMachineLearning is composed of  $201,549$ binary files including $114,737$ malicious files. These files have been gathered from different sources such as VirusShare\footnote{\url{https://virusshare.com/}}, MalShare\footnote{\url{https://malshare.com/}} and TheZoo\footnote{\url{https://github.com/ytisf/theZoo}}. 

\begin{figure}[!ht]
    \centering
    \begin{subfigure}[!ht]{0.4\textwidth}
        \centering
        \includegraphics[width=0.45\textwidth]{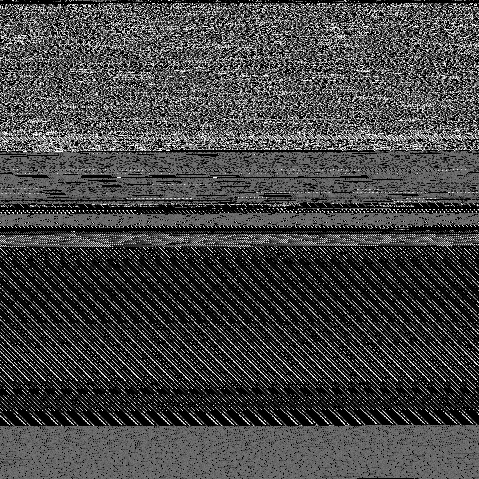}
        \hspace{0.2cm}
        \includegraphics[width=0.45\textwidth]{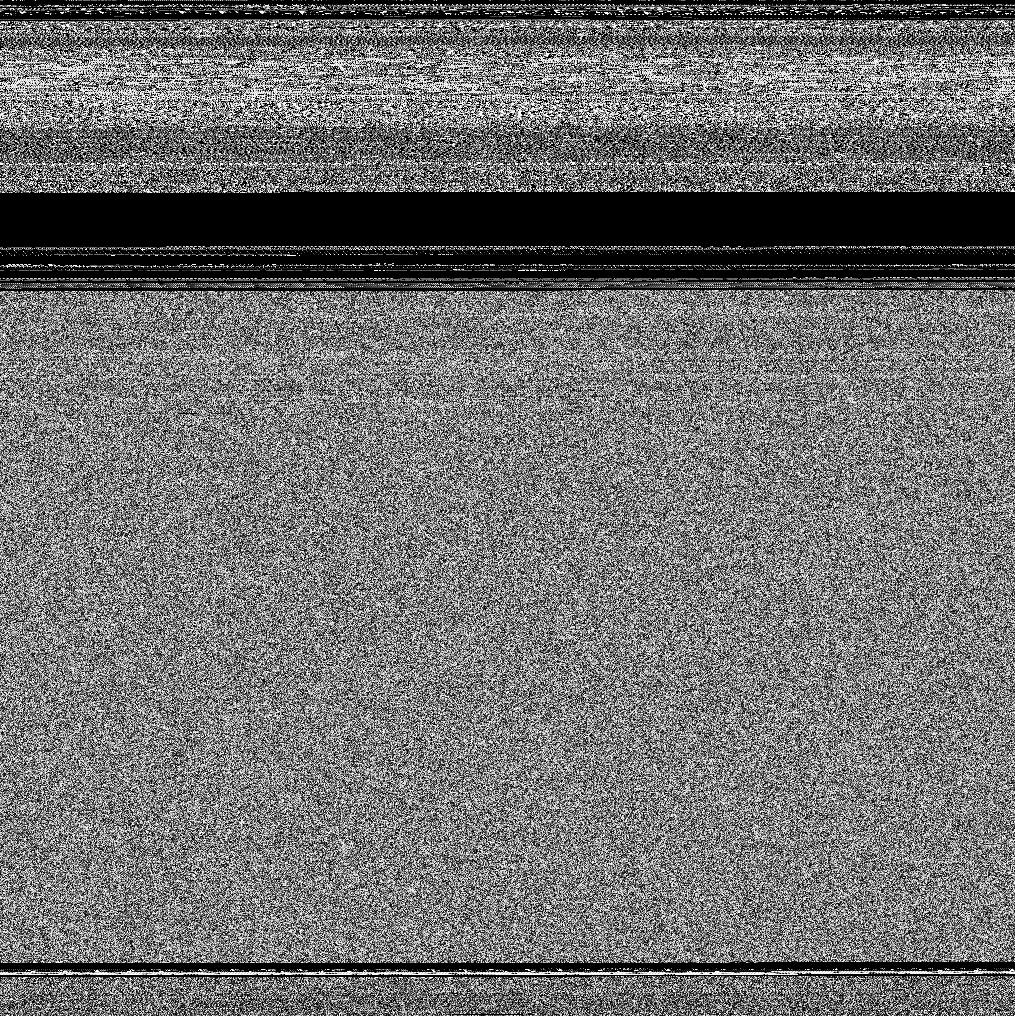}
        \caption{Benign files}
        \label{fig:grayscale_1}
    \end{subfigure}
    % \hspace{0.2cm}
    % \begin{subfigure}[!ht]{0.2\textwidth}
    %     \includegraphics[width=\textwidth]{benign2.png}
    %     \caption{Benign file}
    % \end{subfigure}
    \hspace{0.5cm}
    \begin{subfigure}[!ht]{0.4\textwidth}
        \centering
        \includegraphics[width=0.45\textwidth]{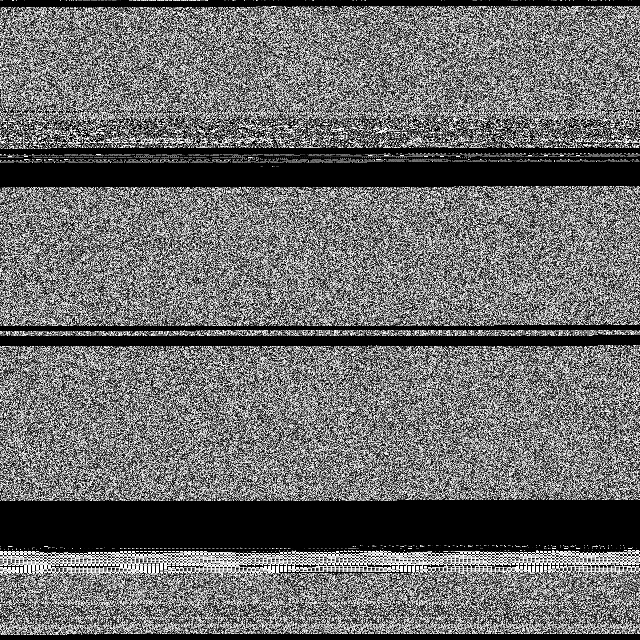}
        \hspace{0.2cm}
        \includegraphics[width=0.45\textwidth]{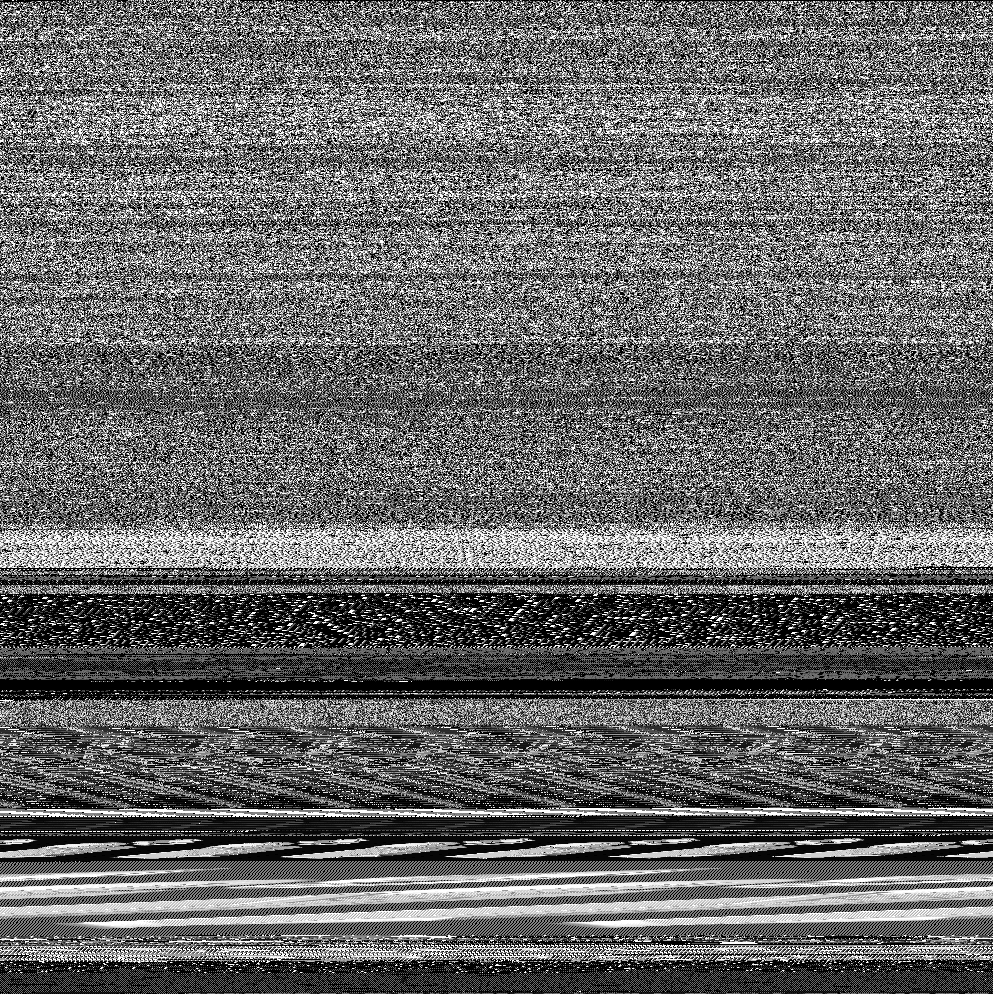}
        \caption{Malicious files}
        \label{fig:grayscale_2}
    \end{subfigure}
    \caption{Examples of PE files transformation into grayscale images. One can see that there is no distinguishable pattern that allows to visually classify malicious and benign files. }
    \label{fig:grayscale}
\end{figure}

The preprocessing that we use in this work is based on the Grayscale method, which  was initially submitted by Nataraj and al. \cite{Nataraj2011} and converts a binary file into an image, as can be seen in Figure \ref{fig:grayscale}. To train our models, we choose to resize Grayscale images to $64\times64$ pixels.

This  dataset may seem unusual or even unsuitable to test the efficiency our new architecture, but it provides a practical application for quantum classification tasks, and Malware detection is a very important research domain in cybersecurity. Besides, this dataset contains more than 300,000 samples, which allows wider tests than most of the datasets that are used for benchmarking tasks. This dataset is also more difficult to classify, as images of malware and benign files are very similar, and we need to use new non-trivial datasets in order to evaluate our models, as suggested in \cite{bermejo2024quantumconvolutionalneuralnetworks}. 

We also used the MNIST dataset  in order to make sure that our algorithms works efficiently and that the method that we propose in this article is well illustrated. We also resize the images from the MNIST dataset to the $64\times64$ size. From this dataset, we first extract only images labeled by `0' or `1', in order to compute a bipartite classification model. Then we also test our algorithm for images labeled by `0' and `8' which are more similar numbers, to challenge the efficiency of our model.  

\section{Description of the QCNN}
\label{sec:descriptionQCNN}

A QCNN is an algorithm composed of an encoding layer, several convolutional and pooling layers, and finally a measurement operator, called observable. Figure \ref{fig:standardQCNN} presents an example of such an algorithm with three layers, where $U_\phi$ corresponds to the encoding layers,  $C_i$ correspond to the gates composing one convolutional layer, and $P_i$ correspond to the gates composing one pooling layer, $i=1,2,3$. In our case the observable is the Z-measurement. 

\begin{figure}[h]
    \centerline{\Qcircuit @C=0.6em @R=.1em {& \multigate{7}{U_\phi} & \qw & \multigate{1}{C_1} & \qw &\ctrl{1} \\
& \ghost{U_\phi} & \qw & \ghost{C_1} & \multigate{1}{C_1}  & \gate{P_1} & \qw & \multigate{2}{C_2} & \qw & \ctrl{2} \\
& \ghost{U_\phi} & \qw & \multigate{1}{C_1} & \ghost{C1}  & \ctrl{1} \\
& \ghost{U_\phi} & \qw & \ghost{C_1} & \multigate{1}{C1}  & \gate{P_1} & \qw & \ghost{C_2} & \multigate{2}{C_2} & \gate{P_2} & \qw & \multigate{4}{C_3} & \qw & \ctrl{4} \\
& \ghost{U_\phi} & \qw & \multigate{1}{C_1} & \ghost{C1}  & \ctrl{1} \\
& \ghost{U_\phi} & \qw & \ghost{C_1} & \multigate{1}{C1}  & \gate{P_1} & \qw & \multigate{2}{C_2} & \ghost{C2} & \ctrl{2} \\
& \ghost{U_\phi} & \qw & \multigate{1}{C_1} & \ghost{C1} & \qw & \ctrl{1} \\
& \ghost{U_\phi} & \qw & \ghost{C_1} &\qw & \gate{P_1} & \qw & \ghost{C_2} & \qw & \gate{P_2} & \qw & \ghost{C_3} & \qw  & \gate{P_3} & \qw & \meter & \cw && &&\lstick{\quad \langle \sigma_Z\rangle}
}}
    \caption{Standard architecture of a quantum convolutional neural network, where convolutional and pooling layers alternate up to the measurement by an observable.}
    \label{fig:standardQCNN}
\end{figure}

Then, one can choose between several architectures for the convolutional layers, as for the pooling layers and the encoding map. Our choices are presented in the following sections (see Figure \ref{fig:pooling_convolution}).

% \[\Qcircuit @C=0.7em @R=.5em {& \multigate{7}{U_\phi} & \qw & \multigate{1}{C_1} & \qw &\ctrl{1} \\
% & \ghost{U_\phi} & \qw & \ghost{C_1} & \qw & \gate{P_1} & \qw & \multigate{2}{C_2} & \qw & \ctrl{2} \\
% & \ghost{U_\phi} & \qw & \multigate{1}{C_1} & \qw & \ctrl{1} \\
% & \ghost{U_\phi} & \qw & \ghost{C_1} & \qw & \gate{P_1} & \qw & \ghost{C_2} & \qw & \gate{P_2} & \qw & \multigate{4}{C_3} & \qw & \ctrl{4} \\
% & \ghost{U_\phi} & \qw & \multigate{1}{C_1} & \qw & \ctrl{1} \\
% & \ghost{U_\phi} & \qw & \ghost{C_1} & \qw & \gate{P_1} & \qw & \multigate{2}{C_2} & \qw & \ctrl{2} \\
% & \ghost{U_\phi} & \qw & \multigate{1}{C_1} & \qw & \ctrl{1} \\
% & \ghost{U_\phi} & \qw & \ghost{C_1} & \qw & \gate{P_1} & \qw & \ghost{C_2} & \qw & \gate{P_2} & \qw & \ghost{C_3} & \qw & \gate{P_3} & \qw & \meter & \cw &\stick{\quad \langle \sigma_Z\rangle}
% } \]
% \\ 

\subsubsection*{Encoding map}
For the encoding layer, we use the same procedure that we used for the previous algorithms \cite{barrue2023quantum}, namely we map all input data $x\in\mathbb{R}^n$ in $[0,\frac{\pi}{2}]^n$, and then we apply the encoding map 
\begin{equation}
\begin{split}
    U_\phi: x\mapsto |\phi(x)\rangle = \bigotimes_{i=1}^n\left(\cos(x_i)|0\rangle +\sin(x_i)|1\rangle\right).
\end{split}
\end{equation}

Note that several encoding maps are possible, some of them being more complex than the one we use. It is also possible to adapt data reuploading \cite{P_rez_Salinas_2020}, in the sense that even if for a QCNN we progressively decrease the number of qubits, a solution would rather be to upload some of the features of the input data in every layer of the neural network. For this approach, one needs to identify what features to re-encode, so a feature importance analysis would be necessary.

\subsubsection*{Convolutional and Pooling layers}

The pooling layers reduce the number of qubits by entangling several qubits together and then tracing out some of them. We choose to use simple pooling gates, and to entangle the qubits pair-wise. The pooling gate that we use is described by the scheme in Figure \ref{fig:pooling}. The convolutional layers entangle qubits and use parameterized gates, that will be optimized during the algorithm. It may seem similar to the pooling layers, but the main difference is that the convolutional layer does not change the number of qubits used in the system. There are multiple possibilities in the choice of gates that one can use for this layer. Here we also choose a simple series of gates, as shown in Figure \ref{fig:convolution}.

\begin{figure}[!h]
    \centering
    \begin{subfigure}[!h]{0.2\textwidth}
        \centerline{\Qcircuit @C=1em @R=.7em {& \ctrl{1} & \qw & \ctrlo{1} & \qw  \\
        & \gate{R_Z(\theta_1)} & \qw & \gate{R_X(\theta_2)} & \qw  }}
        \caption{Pooling layer}
        \label{fig:pooling}
    \end{subfigure}
    \hspace*{3cm}
    \begin{subfigure}[!h]{0.3\textwidth}
        \centerline{\Qcircuit @C=1em @R=.7em { & \gate{R_Y(\theta_1)} &\ctrl{1} & \qw \\
         & \gate{R_Y(\theta_2)} & \targ & \qw }}
        \caption{Convolutional layer}
        \label{fig:convolution}
    
    \end{subfigure}
    \caption{Description of the convolutional and pooling layers used in our algorithm. In the pooling layer, the symbol $\bullet$ means that the gate $R_Z(\theta_1)$ is activated only if the first qubit is in the state $|1\rangle$. Conversely, the symbol $\circ$ means that the gate $R_X(\theta_2)$ is activated only if the first qubit is in the state $|0\rangle$. At the end of the pooling subcircuit we trace out the control qubit in order to reduce the dimension.}
    \label{fig:pooling_convolution}
\end{figure}

 \subsubsection*{Measure}
 At the end of the quantum circuit, we measure the last qubit in the Z-basis in order to obtain the expectation value  $\langle \Psi | \sigma_Z | \Psi\rangle$. In other words, if we denote by $U_\theta$ the matrix describing the quantum circuit, and if we define the observable $\mathcal{O}=Z$, we obtain $$\langle \phi(x)|U_\theta^\dagger\mathcal{O}U_\theta|\phi(x)\rangle.$$ 

Having this expectation value is equivalent to having the probabilities of the quantum state to be in the state $|0\rangle$ or in the state $|1\rangle$. Thus we can use it to compute a cost function, that will be minimized using a gradient-based method. In our case, we use the cross-entropy 
\begin{equation}
    C(\theta)=-\sum_{i=1}^N\left[y_i\log\left(\mathbb{P}(\Psi_i(\theta)=1)\right) + (1-y_i)\log\left(\mathbb{P}(\Psi_i(\theta)=0)\right)\right], 
\end{equation}
where $y_i$ is the label associated to the $i$-th input data $x_i$, and where we measure the probabilities of the state $|\Psi_i(\theta)\rangle = U_\theta|\phi(x_i)\rangle$ to be measured as 0 or 1.  For the gradient-based method, we dispose of several methods, for instance the parameter shift rule \cite{Schuld_2019} and the method called Simultaneous
Perturbation for Stochastic Backpropagation (SPSB) \cite{Hoffmann2022}, that we already used for the two previous tested quantum algorithms. In this article, we only use SPSB, as it is faster than the parameter shift rule and according to our observations \cite{barrue2023quantum} provides comparable results.

\subsection*{Mathematical framework}

We place ourselves in the framework of \cite{Pesah_2021}, to be sure that our algorithm will not suffer from Barren Plateaus. Thus, the gates in the convolutional layers can be defined as gates $W_{i,j}^\ell(\theta^\ell)$ with $$W_{i,j}^\ell(\theta^\ell)=e^{-i\theta^\ell_2H_2}e^{-i\theta^\ell_1H_1}V_{i,j},$$
where $\ell$ represents the layer of the circuit, $\theta^\ell=(\theta_1^\ell, \theta_2^\ell)$, $H_k$ is a Hermitian operator such that $e^{-i\theta^\ell_kH_k}$ is a parametrized single-qubit gate, and $V_{i,j}$ is an unparametrized two-qubit gate (in our case a CNOT gate).

Concerning the pooling layer, the gates in the layers can be thought as maps from a two-qubit Hilbert space to a single-qubit Hilbert space. For a two-qubit state $|\psi\rangle$, the action of the pooling subcircuit is given by $\psi_2=\text{Tr}_1(P_{i,j}^\ell\psi)$, with $\text{Tr}_1$ being the partial trace over the Hilbert subspace corresponding to the first qubit, and where $$P_{i,j}^\ell= |0\rangle\langle 0|_iU_{0,j}^\ell+|1\rangle\langle1|_iU_{1,j}^\ell,$$ where $U_{k,j}^\ell$ are parametrized single-qubit gates acting on the second register of $|\psi\rangle$.
Finally, we measure the Z operator, corresponding to the Pauli-Z gate. We are in a very similar case to the one used in the numerical simulations of \cite{Pesah_2021}, thus the Corollary 1 of \cite{Pesah_2021} can be applied and our quantum circuit will not exhibit Barren plateaus.

%\subsection{Results}

\section{Reuploading circuit}
\label{sec:Reuploading_circuit}

\subsection{Motivations and description of the circuit}

In this section we present a new architecture for the QCNN. Our thinking is that when the number of parameterized gates is too high compared to the number of features, the information given by the encoded data gets lost, thus the performance decreases. We want to use more information, without increasing the number of qubits. Hence, we adapt the concept of data reuploading \cite{P_rez_Salinas_2020}, and reupload more information between each layer of the QCNN. As the number of qubits decreases between two layers, we do not make a choice in the data that we reupload but rather upload new information. For instance, when using a QCNN on 8 qubits (3 layers), we use a PCA 14 on the dataset, then encode the first 8 features before the first layer, the next 4 features between the first and second layers, and the last two features before the third layer. This procedure allows to increase the dimension of the PCA, and thus loose less information about the data: for a $n$-qubits circuit, we are able to use $2(n-1)$ features, whereas the standard procedure only used $n$ features.  

\begin{figure}[h]
    \centerline{\Qcircuit @C=0.62em @R=.45em {& \multigate{7}{U_{\phi_1}}  & \multigate{7}{C_1} & \ctrl{1} \\
& \ghost{U_{\phi_1}}  & \ghost{C_1} & \multigate{6}{P_1} & \qw & \multigate{6}{U_{\phi_2}} & \multigate{6}{C_2} &  \ctrl{2} \\
& \ghost{U_{\phi_1}}  & \ghost{C_1} & \ghost{P_1} \\
& \ghost{U_{\phi_1}}  & \ghost{C_1} & \ghost{P_1} & \qw & \ghost{U_{\phi_2}} & \ghost{C_2} & \multigate{4}{P_2} & \qw & \multigate{4}{U_{\phi_3}}& \multigate{4}{C_3} & \ctrl{4} \\
& \ghost{U_{\phi_1}}  & \ghost{C_1} & \ghost{P_1}  \\
& \ghost{U_{\phi_1}}  & \ghost{C_1} &   \ghost{P_1} & \qw & \ghost{U_{\phi_2}} & \ghost{C_2} &  \ghost{P_2}  \\
& \ghost{U_{\phi_1}}  & \ghost{C_1} & \ghost{P_1}  \\
& \ghost{U_{\phi_1}}  & \ghost{C_1} & \ghost{P_1} & \qw & \ghost{U_{\phi_2}} & \ghost{C_2} &  \ghost{P_2} & \qw &\ghost{U_{\phi_3}}& \ghost{C_3}  & \gate{P_3} & \qw & \meter & \cw & & && \lstick{\quad \langle \sigma_Z\rangle}
}}
    \caption{Description of our architecture for the layered uploading QCNN. We add an encoding layer before each convolutional layer, almost doubling the number of features injected in the quantum circuit.}
    \label{fig:uploading_archi}
\end{figure}

% \[\Qcircuit @C=0.7em @R=.5em {& \multigate{7}{U_{\phi_1}}  & \multigate{7}{C_1} & \ctrl{1} \\
% & \ghost{U_{\phi_1}}  & \ghost{C_1} & \multigate{6}{P_1} & \qw & \multigate{6}{U_{\phi_2}} & \multigate{6}{C_2} &  \ctrl{2} \\
% & \ghost{U_{\phi_1}}  & \ghost{C_1} & \ghost{P_1} \\
% & \ghost{U_{\phi_1}}  & \ghost{C_1} & \ghost{P_1} & \qw & \ghost{U_{\phi_2}} & \ghost{C_2} & \multigate{4}{P_2} & \qw & \multigate{4}{U_{\phi_3}}& \multigate{4}{C_3} & \ctrl{4} \\
% & \ghost{U_{\phi_1}}  & \ghost{C_1} & \ghost{P_1}  \\
% & \ghost{U_{\phi_1}}  & \ghost{C_1} &   \ghost{P_1} & \qw & \ghost{U_{\phi_2}} & \ghost{C_2} &  \ghost{P_2}  \\
% & \ghost{U_{\phi_1}}  & \ghost{C_1} & \ghost{P_1}  \\
% & \ghost{U_{\phi_1}}  & \ghost{C_1} & \ghost{P_1} & \qw & \ghost{U_{\phi_2}} & \ghost{C_2} &  \ghost{P_2} & \qw &\ghost{U_{\phi_3}}& \ghost{C_3}  & \gate{P_3} & \qw & \meter & \cw &\stick{\quad \langle \sigma_Z\rangle}
% } \]
% \\ 

\subsection{Experiments and results}

The main question about this new architecture is to know if it is indeed more efficient than the standard architecture with just one encoding layer at the beginning of the quantum circuit. Thus, our first experiment is to compare the two architectures on the well-known dataset MNIST, in order to classify images representing 0 and 1. We use five epochs for the training of our algorithms, on 10 000 data, gathering the accuracy score for each epoch, and then we gather the accuracy and F1 scores for the tests on 4 000 data. Note that with our chosen architecture, a QCNN with $n$ layers exactly contains $6(2^{n}-1)-2n$ trainable parameters ($2^n$ representing the number of total qubits used in the model), whether we use layered uploading or not. Indeed, since our architecture uses two parameters per convolutional brick and two parameters per pooling brick, the $k$-th  convolutional layer of the model contains $2(2^{n-k+1}-1)$ parameters, and the $k$-th pooling layer contains $2^{n-k+1}$ parameters. When we sum all these layers we obtain a total of $6(2^{n}-1)-2n$ trainable parameters. However, without layered uploading, a $n$-layered QCNN takes only $2^n$ features as input (one feature per qubit), whereas with layered uploading it can use $2(2^n-1)$ features. We compare the two architectures for different numbers of layers, namely from two to four layers. 

We run our tests on Qiskit simulator, for QCNNs models with 2, 3 and 4 layers (thus respectively 4, 8 and 16 qubits), over 5 epochs. For the gradient descent method we use Simultaneous Perturbation for Stochastic Backpropagation \cite{Hoffmann2022}, with a learning rate of $0.1$.

\begin{table}[htp]
    \centering
    \small
    \resizebox{\columnwidth}{!}{
    \begin{tabular}{|c|c|c|c|c|c|c|c|} 
       \hline
       layers & uploading & Metric & e1 & e2 & e3 & e4 & e5  \\
       \hline
       \multirow{6}{*}{2 layers}  & \multirow{3}{*}{With} & train acc & 0.9642 & 0.977 & 0.9801 & 0.9825  & 0.9827  \\ \cline{3-8} 
       & & test acc & 0.9917 &0.9712 &0.9742 & 0.9847 & 0.9877 \\ \cline{3-8} 
       & & F1-Score &0.9922 & 0.9735 & 0.9762 & 0.9858 &0.9885 \\ \cline{2-8}
            & \multirow{3}{*}{Without} & train acc & 0.9606 & 0.9849 & 0.9854 & 0.9811 & 0.9829  \\ \cline{3-8} 
       & & test acc & 0.98 & 0.9732 & 0.9832 &0.9705 & 0.9812\\ \cline{3-8} 
       & & F1-Score & 0.9814 & 0.9753 & 0.9844 & 0.9729 & 0.9826 \\ 
       \Xhline{3\arrayrulewidth}
       \multirow{6}{*}{3 layers}  & \multirow{3}{*}{With} & train acc & 0.9724 & 0.9855 & 0.9852 & 0.9825  &  0.9864 \\ \cline{3-8} 
       & & test acc &0.9787 & 0.9725 & 0.9707 & 0.9800& 0.9855\\ \cline{3-8} 
       & & F1-Score & 0.9803 &0.9747 & 0.9731 & 0.9814 & 0.9864 \\ \cline{2-8}
            & \multirow{3}{*}{Without} & train acc &0.9739 & 0.9807 & 0.9865 & 0.9881 & 0.9870  \\ \cline{3-8} 
       & & test acc & 0.9850 & 0.9640 &0.9750 & 0.9895 & 0.9850\\ \cline{3-8} 
       & & F1-Score & 0.9860 &0.9671 &0.9769 & 0.9901 & 0.9860 \\ 
       \Xhline{3\arrayrulewidth}
       \multirow{6}{*}{4 layers}  & \multirow{3}{*}{With} & train acc & 0.9508 & 0.9869 & 0.9874 & 0.9873  & 0.9880  \\ \cline{3-8} 
       & & test acc&0.9827 & 0.9775 & 0.9875 & 0.9902 & 0.9867 \\ \cline{3-8} 
       & & F1-Score & 0.9839 & 0.9791 & 0.9883 & 0.9908 & 0.9876\\ \cline{2-8}
            & \multirow{3}{*}{Without} & train acc&0.8766  & 0.9770 & 0.9823 &0.9837 & 0.9839 \\ \cline{3-8} 
       & & test acc& 0.9647 & 0.9665 & 0.9897 &0.9775 &0.9592 \\ \cline{3-8} 
       & & F1-Score & 0.9678 & 0.9693 & 0.9904 & 0.9791 & 0.9629 \\ 
       \hline
    \end{tabular}}
    \normalsize
    \caption{Comparison of the two architectures for different layers on the MNIST dataset, classifying images of 0 and 1. For each model we gather the train  and test accuracy and the F1-score after every epoch.}
    \label{tab: MNIST0vs1}
\end{table}
% \begin{table}[h]
%     \centering
%     \small
%     \resizebox{\columnwidth}{!}{
%     \begin{tabular}{|c|c|c|c|c|c|c|c|} 
%        \hline
%        layers & uploading & e1 & e2 & e3 & e4 & e5 & test(acc/F1) \\
%        \hline
%        \multirow{2}{*}{2 layers}  & with & 0.9695 & 0.98 & 0.9819 & 0.9774 & 0.9808 & 0.9837/0.9848 \\ \cline{2-8}
%             & without & 0.9549 & 0.9795 & 0.9786 & 0.9787 & 0.9783 & 0.9742/0.9762 \\ 
%         \hline 
%         \multirow{2}{*}{3 layers}  & with & 0.9292 & 0.9701 & 0.9645 & 0.97 & 0.9652 & 0.983/0.9841 \\ \cline{2-8}
%             & without & 0.8908 & 0.968 & 0.9661 & 0.9757 & 0.9685 & 0.97275/0.9746 \\
%         \hline
%         \multirow{2}{*}{4 layers}  & with & 0.7102 & 0.9558 & 0.96 & 0.9607 & 0.9577 & 0.9795/0.9809 \\ \cline{2-8}
%             & without & 0.7821 & 0.96 & 0.9709 & 0.9737 & 0.9706 & 0.97725/0.9789\\
%         \hline
%     \end{tabular}}
%     \normalsize
%     \caption{Comparison of the two architectures for different layers on the MNIST dataset, classifying images of 0 and 1. }
%     \label{tab:MNIST0vs1}
% \end{table}

As we can see, the new architecture  performs better in every model, as in terms of accuracy or F1 scores. If we look at the training performances, sometimes the model without uploading gets better results, but since in the test it is not the case it could suggest some overfitting phenomenon. Note also that the model performances seems to decrease with the number of layers, which tends to confirm our intuition that the information from the input data may get lost when there are too many parameters in the quantum circuit. With this table one can compare a model with and without layered uploading to see that the new architecture achieves better results, which is quite logical because more features were added in the model. On the other hand, one can also compare a $n$-layers model with layered uploading, to a $(n+1)$-layers model without layered uploading. In this case, the two models have very similar numbers of features (for instance a 3-layers model with layered uploading uses 14 features while a 4-layers model uses 16 features), and differ by their size: the $(n+1)$-layers model is encoded on $2^n$ more qubits and contains $3\times2^{n+1}-2$ more trainable parameters. Results show that in each case the $n$-layers model performs better, the gap between the two models increasing when the number of layers $n$ increases. All these previous remarks concern the models looking at their results in the final epochs. One can also wonder which architecture is the best when dealing with only one or two epochs. In this case it is less clear but our layered uploading method also seems to outperform the standard version in most of the cases. 

These results  show that our architecture increases the performances of the QCNN model. However, the difference between the scores of two architectures is quite thin, and we would like to see if for more complex cases it becomes more significant. Thus we try a second test on the MNIST dataset. This time we want to classify images corresponding to 0 ad 8, which are graphically closer numbers than 0 and 1.   

\begin{table}[htp]
    \centering
    \small
    \resizebox{\columnwidth}{!}{
    \begin{tabular}{|c|c|c|c|c|c|c|c|} 
       \hline
       layers & uploading & Metric & e1 & e2 & e3 & e4 & e5  \\
       \hline
       \multirow{6}{*}{2 layers}  & \multirow{3}{*}{With} & train acc & 0.8934 & 0.9459 & 0.9506 & 0.9468 &  0.9395 \\ \cline{3-8} 
       & & test acc & 0.9515 &0.9177 & 0.952 &0.9375 & 0.9482 \\ \cline{3-8} 
       & & F1-Score & 0.9517& 0.9210 & 0.9517 & 0.9386 & 0.9483\\ \cline{2-8}
            & \multirow{3}{*}{Without} & train acc & 0.9463 & 0.9587  & 0.9566 & 0.9627 &  0.9588 \\ \cline{3-8} 
       & & test acc &0.9622 & 0.9572 & 0.9672&0.9150 & 0.9152 \\ \cline{3-8} 
       & & F1-Score &0.9615 & 0.9573 & 0.9664 & 0.9191&0.9188 \\ 
       \Xhline{3\arrayrulewidth}
       \multirow{6}{*}{3 layers}  & \multirow{3}{*}{With} & train acc& 0.9222 & 0.9418 & 0.9406 & 0.9394 &  0.9443 \\ \cline{3-8} 
       & & test acc & 0.9405 & 0.946 &0.9362 & 0.9407 & 0.94925 \\ \cline{3-8} 
       & & F1-Score & 0.9412 & 0.9460 & 0.9368 & 0.9412 & 0.9486\\ \cline{2-8}
            & \multirow{3}{*}{Without} & train acc & 0.9141 & 0.9405  & 0.9436 & 0.9476 &  0.9402 \\ \cline{3-8} 
       & & test acc & 0.9402& 0.9120 & 0.94 & 0.9385 & 0.9475 \\ \cline{3-8} 
       & & F1-Score & 0.9411 &0.9159 &0.9407 &0.9394 & 0.9470 \\ 
       \Xhline{3\arrayrulewidth}
       \multirow{6}{*}{4 layers}  & \multirow{3}{*}{With} & train acc & 0.8412 & 0.9448 & 0.9347 & 0.9289 &  0.9404 \\ \cline{3-8} 
       & & test acc& 0.946 & 0.8995 &0.9497 & 0.942& 0.9252 \\ \cline{3-8} 
       & & F1-Score &0.9465 &0.9049 & 0.9501 &0.9424 & 0.9276 \\ \cline{2-8}
            & \multirow{3}{*}{Without} & train acc & 0.8718 & 0.9430  & 0.9420 & 0.9404 & 0.9421  \\ \cline{3-8} 
       & & test acc &0.9435 & 0.9365 & 0.9215 &0.9350 & 0.91925\\ \cline{3-8} 
       & & F1-Score &0.9424 &0.9374 &0.9243 &0.9362 & 0.9224 \\ 
       \hline
    \end{tabular}}
    \normalsize
    \caption{Comparison of the two architectures for different layers on the MNIST dataset, classifying images of 0 and 8. For each model we gather the train  and test accuracy and the F1-score after every epoch.}
    \label{tab:MNIST0vs8}
\end{table}

% \begin{table}[h]
%     \centering
%     \small
%     \resizebox{\columnwidth}{!}{
%     \begin{tabular}{|c|c|c|c|c|c|c|c|}
%        \hline
%        layers & uploading & e1 & e2 & e3 & e4 & e5 & test(acc/F1) \\
%        \hline
%        \multirow{2}{*}{2 layers}  & with & 0.945 & 0.938 & 0.953 & 0.965 & 0.963 & 0.9547/0.9548 \\ \cline{2-8}
%             & without & 0.946 & 0.960 & 0.934 & 0.961 & 0.965 & 0.9512/0.9518 \\ 
%         \hline 
%         \multirow{2}{*}{3 layers}  & with & 0.8992 & 0.9548 & 0.9578 & 0.947 & 0.9499 & 0.958/0.9579 \\ \cline{2-8}
%             & without & 0.9138 & 0.9544 & 0.9499 & 0.9469 & 0.9328 & 0.9495/0.9500 \\
%         \hline
%         \multirow{2}{*}{4 layers}  & with & 0.5265 & 0.5198 & 0.7798 & 0.9454 & 0.9454 & 0.926/0.928 \\ \cline{2-8}
%             & without & 0.8264 & 0.9517 & 0.9447 & 0.9596 & 0.9530 & 0.944/0.9449\\
%         \hline
%     \end{tabular}}
%     \normalsize
%     \caption{Comparison of the two architectures for different layers on the MNIST dataset, classifying images of 0 and 8. }
%     \label{tab:MNIST0vs8}
% \end{table}

On this table we can see the same phenomenon, namely that the model with the layered uploading performs better than the model without layered uploading. The differences between the scores of the two models do not seem to be greater than for the previous test, the gain being around 1\% for both tasks, which is not negligible for this kind of datasets. We also observe that when comparing models with similar numbers of features, the smaller model (namely the one with less layers and layered uploading) outperforms the bigger one, the gap between them increasing as the difference of added trainable parameters increases. For this second classification task, we note that the scores are smaller than when we classify 0 and 1; it can be explained quite simply by the fact that 0 and 8 are graphically closer than 0 and 1, especially with a low dimension of PCA which tends to blur the images. When looking at the results for the first epochs, we cannot really conclude to a better performance of our model, which does better in some cases but not all of them.  

Finally, we want to test the efficiency of our architecture on the dataset which is important to us, namely the malware images dataset. Since this dataset is made of images of files, which are very noisy (see Figure \ref{fig:grayscale}), the classification task seems even harder than previously. The dimension of the PCA thus should play a very important role, and we want it as high as possible without increasing the number of qubits too much. To deal with this dataset, we compute the F1-score for both training and test, because it is a more robust metric when it comes to unbalanced datasets. %We also compute the accuracy scores, which are postponed in the Appendix.  

\begin{table}[htp]
    \centering
    \small
    \resizebox{\columnwidth}{!}{
    \begin{tabular}{|c|c|c|c|c|c|c|c|} 
       \hline
       layers & uploading & Metric & e1 & e2 & e3 & e4 & e5  \\
       \hline
       \multirow{6}{*}{2 layers}  & \multirow{3}{*}{With} & \footnotesize train F1-S  & 0.6680 & 0.6927 & 0.6731 & 0.6736 & 0.6769 \\ \cline{3-8} 
       & & test acc & 0.7202 & 0.71 & 0.7392 & 0.7135 & 0.735 \\ \cline{3-8} 
       & & test F1-S & 0.7588 & 0.6974 & 0.7369 & 0.7326 & 0.7613 \\ \cline{2-8}
            & \multirow{3}{*}{Without} & \footnotesize train F1-S  & 0.6823 & 0.6735 & 0.7137 & 0.6816 & 0.6718  \\ \cline{3-8} 
       & & test acc & 0.6995 & 0.6622 & 0.739 & 0.728 & 0.7432 \\ \cline{3-8} 
       & & test F1-S & 0.6924 & 0.6299 & 0.7709 & 0.7274 & 0.7596 \\ 
       \Xhline{3\arrayrulewidth}
       \multirow{6}{*}{3 layers}  & \multirow{3}{*}{With} & \footnotesize train F1-S  & 0.6173 & 0.6382 & 0.6702 & 0.6638 & 0.6997 \\ \cline{3-8} 
       & & test acc & 0.702 & 0.7172 & 0.7065 & 0.6795 & 0.7157 \\ \cline{3-8} 
       & & test F1-S  & 0.7345 & 0.7389 & 0.7385 & 0.7276 & 0.7452 \\ \cline{2-8}
            & \multirow{3}{*}{Without} & \footnotesize train F1-S  & 0.6285 & 0.6606 & 0.6680 & 0.6258 & 0.6381 \\ \cline{3-8} 
       & & test acc & 0.6592 & 0.6165 & 0.6245 & 0.5582 & 0.6910 \\ \cline{3-8} 
       & & test F1-S & 0.6912 & 0.7116 & 0.5403 & 0.6860 & 0.7385 \\ 
       \Xhline{3\arrayrulewidth}
       \multirow{6}{*}{4 layers}  & \multirow{3}{*}{With} & \footnotesize train F1-S & 0.5614 & 0.5742 & 0.5061 & 0.5646 & 0.6208 \\ \cline{3-8} 
       & & test acc & 0.672 & 0.502 & 0.5182 & 0.5167 & 0.5202 \\ \cline{3-8} 
       & & test F1-S  & 0.6543 & 0.4292 & 0.2897 & 0.6687 & 0.6630 \\ \cline{2-8}
            & \multirow{3}{*}{Without} & \footnotesize train F1-S & 0.5236 & 0.4879 & 0.5748 & 0.6611 & 0.4715 \\ \cline{3-8} 
       & & test acc & 0.505 & 0.525 & 0.5735 & 0.5235 & 0.5392 \\ \cline{3-8} 
       & & test F1-S & 0.6587 & 0.1724 & 0.3809 & 0.6550 & 0.6177\\ 
       \hline
    \end{tabular}}
    \normalsize
    \caption{Comparison of the two architectures for different layers on the malware dataset, classifying images of malware and benign files. For each model and each epoch, we compute the train and test F1-Score, and the test accuracy.}
    \label{tab:malware}
\end{table}

For the models in Table \ref{tab:malware}, we observe once again that circuit with layered uploading performs better on the F1-score: around 1\% for two layers, and a lot more for the three-layers model. Concerning the four-layers model, we can see by looking the test accuracy results that the model struggles to learn, and this is highlighted by the values of the test F1-score in the standard model.   This might be due to the number of parameters which is too big compared to the number of features injected in the circuit, which themselves are maybe be not meaningful enough. When comparing a $n$-layers model with layered uploading to a $(n+1)$-layers model without layered uploading (in order to deal with similar numbers of features), we can see that in every case the $n$-layers model achieves better F1-Scores on the test in the last epochs. It is less clear on the first epochs, which may be due for example to the randomness of the parameters initialization.  These results give a very good example of the utility of our architecture, which allows to have a good learning phase even with three layers and low-impact features. For the bigger model, we can see however that both of the algorithms perform quite poorly in the detection task, reaching less than 67\% in F1-score. Thus, even if there is a difference of 5\% between the two architectures we can consider that the added features are not impactful enough to keep good performances, and that the number of free parameters is too big compared to the number of features, so that the information about the files might get lost in the quantum circuit. Besides, the PCA may not be a suitable solution for the feature reduction, because it does not give us any information on the relevance of these new features. Moreover, in our use-case, we classify images of PE files, composed of different specific sections, with specific features, which are mixed up all together when applying the PCA. A deeper analysis of the structure of our dataset with standard QCNNs can be found in  \cite{quertier2023indepth}. 

When the size of the quantum circuit increases, initialization of parameters becomes important. Even if in our case we do not exhibit Barren plateaus, a bad initialization could slow down the learning phase, and thus give poor results. We try to initialize all our parameters to 0, and then to $2\pi$, in order to start the training with the identity operator, but it did not increase the model's performances. To reduce this issue, a good solution could be to run our experiments with several sets of random initial parameters, and to average the results in order to get a more stable estimation of the performances. We tried this approach for simple models only, due to computational cost and time issues.  We ran our model with two layers and layered uploading on five random initializations of parameters, and compute the mean of the performances of these five tests. Then we did the same thing for the model with two layers without layered uploading (and the same initialization parameters), and for the model with three layers without layered uploading, in order to make the same comparisons as before (in this case, we have to take different initializations because the 3-layers model has more parameters). We tested this method on both MNIST01 (classification of 0 and 1) and on the Malware dataset.  

\begin{figure}[!ht]
    \centering
    \begin{subfigure}[!ht]{0.8\textwidth}
        \centering
        \includegraphics[width=\textwidth]{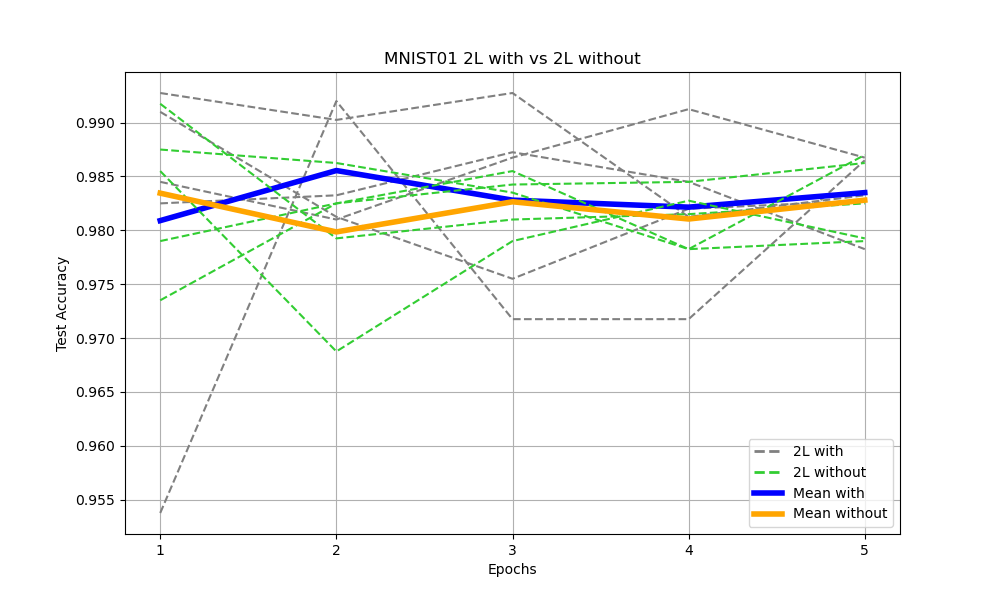}
        \caption{2 layers with uploading against 2 layers without uploading}
    \end{subfigure}
    \hspace{0.5cm}
    \begin{subfigure}[!ht]{0.8\textwidth}
        \centering
        \includegraphics[width=\textwidth]{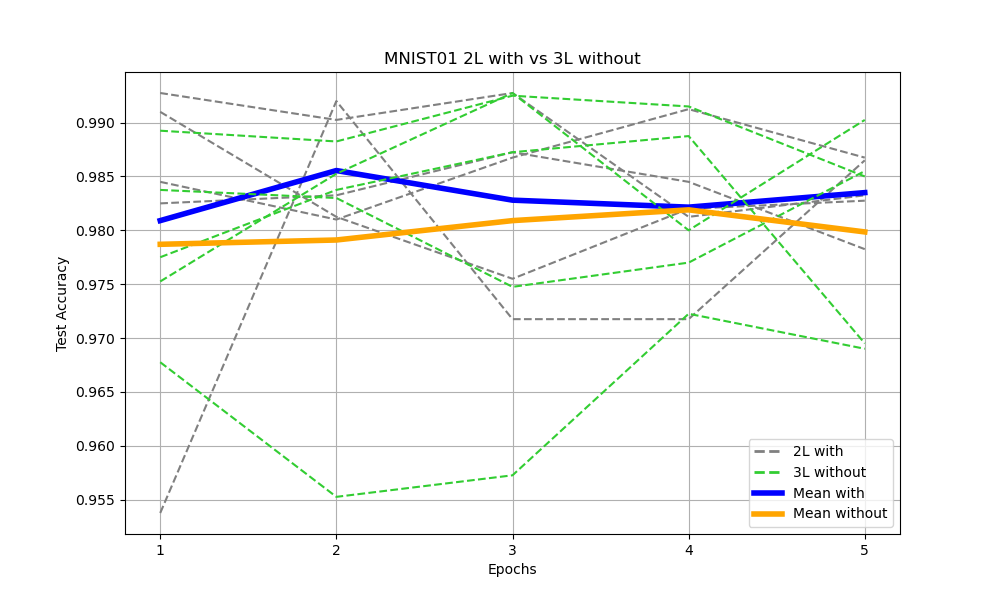}
        \caption{2 layers with uploading against 3 layers without uploading}
    \end{subfigure}
    \caption{Comparison of 5 random initializations of parameters for models with uploading and 2 layers, without uploading and 2 layers, and without uploading and 3 layers, on the MNIST dataset. For the models with 2 layers we use the same initialization parameters. We also plot the average of the 5 initializations, showing that the model with layered uploading is indeed better than the models without layered uploading.}
    \label{fig:comparison_MNIST}
\end{figure}

\begin{figure}[!ht]
    \centering
    \begin{subfigure}[!ht]{0.8\textwidth}
        \centering
        \includegraphics[width=\textwidth]{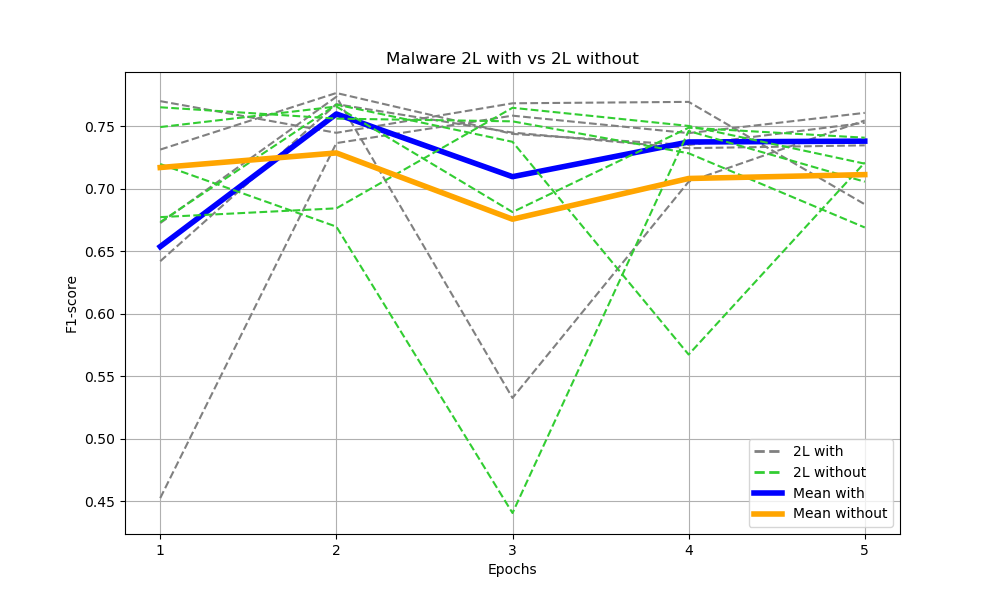}
        \caption{2 layers with uploading against 2 layers without uploading}
    \end{subfigure}
    \hspace{0.5cm}
    \begin{subfigure}[!ht]{0.8\textwidth}
        \centering
        \includegraphics[width=\textwidth]{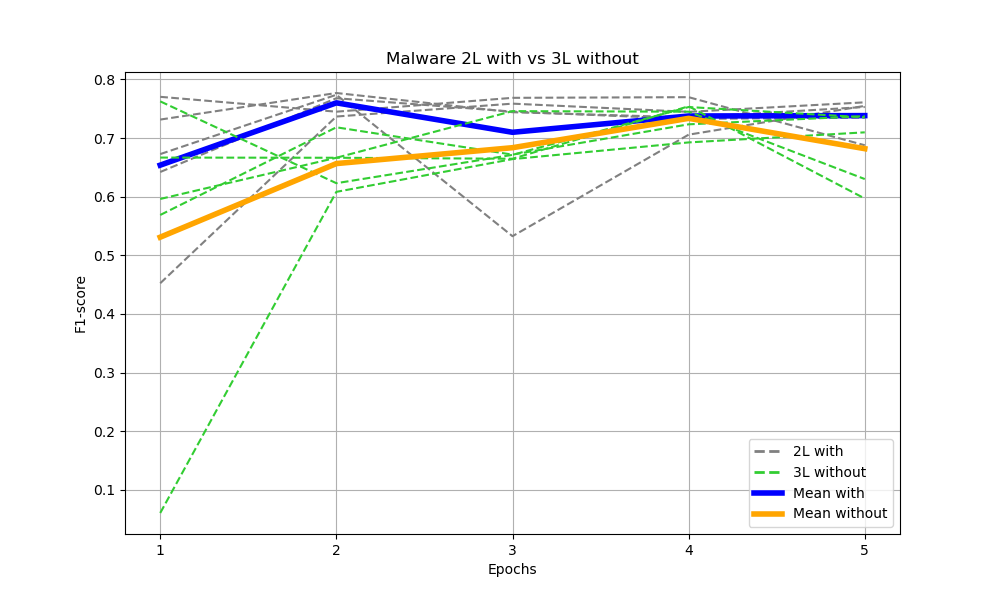}
        \caption{2 layers with uploading against 3 layers without uploading}
    \end{subfigure}
    \caption{Comparison of 5 random initializations of parameters for models with uploading and 2 layers, without uploading and 2 layers, and without uploading and 3 layers, on the Malware dataset. For the models with 2 layers we use the same initialization parameters. We also plot the average of the 5 initializations, showing that the model with layered uploading is indeed better than the models without layered uploading.}
    \label{fig:comparison_MNalware}
\end{figure}

Figure \ref{fig:comparison_MNIST} shows the results for the MNIST dataset, while Figure \ref{fig:comparison_MNalware} shows the results on the Malware dataset. For each experiment, we can see that on average the model with layered uploading performs better that both models without layered uploading, which is particularly convenient when we compare it to the 3-layers model because it can prevent us to use a bigger model. Note that the differences between the models are more important in the Malware dataset than in the MNIST dataset. It is due to the fact that all models have already great accuracy on this last dataset, so the performances are more similar. This also confirms the suitability of our dataset for some benchmarking tasks.

Tables \ref{tab:mean_accuracy} and \ref{tab:mean_accuracy_malware} show respectively the mean accuracy for the three models on MNIST and the mean F1-score for the three models on the Malware dataset. Note that in this last case our architecture allows to gain up to 5\% in F1-score, without increasing the number of layers and thus the number of trainable parameters. 

\begin{table}[htp]
    \centering
    \begin{tabular}{|c|c|c|c|c|c|}
    \hline
      Layers & e1 & e2& e3& e4& e5 \\
      \hline
    2L with  & 0.9808& 0.9855 & 0.9827& 0.9821& 0.9834 \\
    \hline
     2L without & 0.98345 & 0.9798 & 0.9826 & 0.9810 & 0.98279 \\
    \hline
     3L without & 0.9787 & 0.9791 & 0.9808 & 0.9818 & 0.9798 \\
    \hline
    \end{tabular}
    \caption{Mean test accuracies - MNIST }
    \label{tab:mean_accuracy}
\end{table}

\begin{table}[htp]
    \centering
    \begin{tabular}{|c|c|c|c|c|c|}
    \hline
      Layers & e1 & e2& e3& e4& e5 \\
      \hline
    2L with & 0.6537 & 0.7599 & 0.7097 & 0.7375& 0.7382
 \\
    \hline
     2L without & 0.7171& 0.7287 & 0.6757& 0.7083& 0.7114 \\
    \hline
     3L without & 0.5309& 0.6564& 0.6835& 0.7335& 0.6818 \\
    \hline
    \end{tabular}
    \caption{Mean F1-scores - Malware }
    \label{tab:mean_accuracy_malware}
\end{table}

%Finally, as another perspective, the size of the dataset could also be increased to see if it makes a difference. However, our aim being to be able to extract useful information for little datasets, with few features, we did not explore these leads, especially because of the computation time problematic, and also because our architecture gave good results for all the tested models but this last one. 

\section*{Conclusion}

In a  nutshell, we proposed in this paper a new architecture for quantum convolutional neural networks, allowing to encode more features without increasing the size of the quantum circuit. It is inspired from data reuploading and consists in adding an encoding layer composed of new features before each convolutional layer. With this approach, one can use $2(n-1)$ features for a $n$-qubits circuit, so it almost doubles the information that is given to the QCNN. We compared our architecture with a standard QCNN on several classification tasks, namely classifying 0 and 1 then 0 and 8 from the MNIST dataset, and classifying malware and benign files from our personal dataset. On all our tests  this new architecture presented better performances than the standard one, so we believe it is a relevant method to use the information that we can gather from data without systematically increase the size of our dataset or the size of the algorithm. This work might for instance allow us to enhance our models using QCNNs as presented in \cite{quertier2023indepth} in order to get a better detection rate and a greater knowledge about malware PE files.  

There are several leads that could be investigated in order to continue our work on this architecture. First we did not focus on the convolutional and pooling layers: maybe some other choices of gates would allow better performances for our use-cases. Computing the Lie algebra generated by each layer could give us information about the expressiveness of our model, as it has been shown to be an important tool for QNNs \cite{Larocca_2021,ragone2023unified,Meyer2022}. Concerning the encoding of the features itself, we think that our limits mainly came  from the fact that the PCA method does not give information about the output features, so it is hard to know if we encode the features in the right way in the quantum circuit. One solution would be to identify in a first time the most important features, and then reorganize the way that we encode them in the circuit. Besides, if some features are largely more impacting than others, one could try to re-encode them all along the algorithm (for example in each encoding layer). Finally, even if it is not in our concern yet because of the current capacities of quantum computers, it would be good to pursue our tests on bigger models an bigger datasets to see if it makes a real difference in the learning of QCNNs.  

%In \cite{Larocca_2021} the authors explain that the Lie algebra generated by these gates can give us information about the overparameterization regime of our model, so it could be interesting to adapt this theory for QCNNs. 

\bibliographystyle{unsrt}
\bibliography{biblio}

\section*{Funding}

The authors declare that there is no funding for this work.

\end{document}